\documentclass[12pt,a4paper]{article}
\usepackage{epsfig}

\def\ts2f{${\cal T}^{*(2F)}$}
\def\tst2f{{\cal T}^{*(2F)}}
\def\t2f{${\cal T}^{(2F)}$}
\def\tsa4{${\cal T}^{*(A_4)}$}
\def\tms{${\cal T}^{(MS)}$}

\def\ep{${{\mbox{I}}\!{\mbox{{\bf E}$_\parallel$ } }}$}
\def\es{${{\mbox{I}}\!{\mbox{{\bf E}$_\perp $} }}$}
\newcommand{\fracds}[2]{\displaystyle \frac{#1}{#2} }
\def\enb{\setlength{\unitlength}{2072sp}%
\begin{picture}(249,249)(1114,-973)
\thinlines
\put(1126,-961){\framebox(225,225){}}
\end{picture}}

\def\ku{l\!\!\! Q}

\def\zfo{${{\mbox{$\bigcirc$}}\!\!\!\!\!\!\:{\mbox{2}}\,}$}
\def\ffo{${{\mbox{$\bigcirc$}}\!\!\!\!\!\!\:{\mbox{5}}\,}$}

\def\ba{\begin{eqnarray}}\def\ea{\end{eqnarray}}
\def\lb{\label}\def\ni{\noindent}
\def\cl{\centerline}
\def\be{\begin{equation}}\def\ee{\end{equation}}

\def\s{\sigma}\def\a{\alpha}\def\b{\beta}\def\g{\gamma}
\def\d{\delta}
\def\t{\tau}\def\l{\lambda}\def\D{\Delta}

\def\ti{\tilde}

\begin{document}

\hfill{\normalsize\bf CPT--99/P.3879}\\
\vskip1.5cm
\cl{\LARGE On Quasiperiodic Space Tilings, Inflation}

\vskip .8cm

\cl{\LARGE and Dehn Invariants }

\vskip1.2cm

\cl{\large Oleg Ogievetsky\footnote{On leave of absence from
P. N. Lebedev Physical Institute,
Theoretical Department, Leninsky pr. 53, 117924 Moscow, Russia}}
\vskip .3cm
\cl{\small\em Center of Theoretical Physics, Luminy,
13288 Marseille, France}
\vskip .15cm
\cl{\small\em e-mail: oleg@cpt.univ-mrs.fr}
\vskip.5cm
\cl{\large and}
\vskip.5cm
\cl{\large Zorka Papadopolos}
\vskip .3cm
\cl{\small\em Institut f\"ur Theoretische Physik, Universit\"at
    Magdeburg, PSF\ 4120,}
\cl{\small\em 39016\ Magdeburg, Germany}
\vskip .15cm
\cl{\small\em e-mail: zorka.papadopolos@uni-tuebingen.de}

\vskip1.6cm

\cl{\bf Abstract}
\medskip

\ni
{\small 
We introduce Dehn invariants as a useful tool in
the study of the inflation of quasiperiodic space tilings.
The tilings by ``golden tetrahedra'' are considered.
We discuss how the Dehn invariants can be applied to the study
of inflation properties of the six golden tetrahedra.
We also use geometry of the faces of the golden tetrahedra
to analyze their inflation properties. We give the 
inflation rules for decorated Mosseri--Sadoc tiles in the  
projection class of tilings ${\cal T}^{(MS)}$.
The Dehn invariants of the Mosseri--Sadoc tiles provide
two eigenvectors of the inflation matrix with eigenvalues
equal to  $\tau = \frac{1+\sqrt{5}}{2}$ and $-\frac{1}{\tau}$,
and allow to reconstruct the inflation matrix uniquely.
}

\newpage

\section{Introduction}

\vskip .6cm
Existing mathematical models of quasiperiodic
tilings of the plane~\cite{P,GS,BKSZ} and
the 3dimensional space~\cite{K,SS,MS,D,PHK}
admit an important operation called inflation.
Given a tiling of the plane or the space by  
prototiles $\{ X^i \}$
from a {\em local isomorphism class of tilings}~\cite{BKSZ}
or {\em specie}~\cite{D},
the inflation produces another tiling of the class out
of the first one, by blowing up the tiles with a factor $\l$
($\l$ is bigger then 1 and called the inflation  factor) and
substituting the $\l$--scaled tiles  $X^i_{(\l )}$ in a particular
way by the tiles $\{ X^i \}$ of the original size.
Generally, in the process of inflation, the tiles
$X^i$ are cut into
pieces (by plane cuts) and these smaller pieces can then be
recombined together into the tiles $X^i_{(\l )}$.
The tile $X^i_{(\l )}$ is made out of pieces of tiles
$X^j$ for all $j$.
Let $M^i_j$ be the sum of volumes of the pieces of the tiles
of the type $X^j$. The matrix $M=M^i_j$ is called the volume
inflation matrix.

By its definition, the matrix $M$ has an eigenvector
$\vec{v}$ with components $v^i={\rm Vol}(X^i)$,
the volumes of the tiles. The corresponding eigenvalue 
is $\l^3$.

In some cases, the matrix $M$ has rational entries. 
An example of an exception is the volume inflation 
matrix of the class of the tilings
\ts2f\ icosahedrally projected from the lattice
$D_6$,~\cite{PHK, HKP}, to be discussed later in this  paper.
(Note: Under the ``icosahedral projection" we mean the
icosahedrally invariant projection.) Let $\ku [\l ]$
be the extension of $\ku$ by $\l$ and
$G={\rm Gal}( \ku [\l ]/ \ku )$ its Galois group. Let
$G\l =\{ \l_1=\l ,\l_2 ,\dots ,\l_k\} $ be the orbit of $\l$.
Then all the $\l_i$ are eigenvalues of the matrix $M$.

In many physically interesting cases, $\l$ is a power of the
golden mean $\t =\frac{1+\sqrt{5}}{2}$; the field $\ku [\l ]$
is quadratic and therefore volumes can be used to build two
eigenvectors (and eigenvalues) of $M$.

In this article we address a question of a geometrical meaning
of other eigenvectors of $M$.

We need several standard definitions. One says that two
polyhedra, $P_1$ and $P_2$, are scissor--equivalent
(notation:  $P_1 \sim P_2$) if $P_1$ can be cut 
(by plane cuts)
and rebuilt into $P_2$.

Assume that there is a function ${\cal F}$ which associates
an element of a ring ${\cal K}$ to any polyhedron.
The function ${\cal F}$ is called scissor--invariant
if ${\cal F}$ enjoys the property: $P_1 \sim P_2$
$\Rightarrow$ ${\cal F}(P_1)={\cal F}(P_2)$.

Any scissor--invariant function ${\cal F}$ allows to construct
an eigenvector $\vec{f}$ of the matrix $M$,
$f^i={\cal F}(X^i)$. The comment about the Galois group
holds for the vector $\vec{f}$ as well.

It is well known that starting from the dimension 3, the
space of scissor--invariant functions is nontrivial:
in addition to the volume, there are also Dehn invariants.

In Section 2 of the present article we remind some basic
facts about the Dehn invariants. 

In Section 3 we consider the Dehn invariants of
golden tetrahedra. We use the Dehn invariants as a
test of an existence of a stone inflation for the
golden tetrahedra (Subsection 3.2). We show that if
a rational inflation (that is, an inflation whose
inflation matrix has rational entries) for the golden tetrahedra
with the inflation factor $\tau$ exists then the inflation matrix
can be uniquely reconstructed with the help of the volumes and 
the Dehn invariants. This unique inflation matrix $M_{gt}$
turns out to have non-integer entries which shows that a stone
inflation of the golden tetrahedra with the inflation factor $\tau$
cannot exist. However, $M_{gt}^3$, the cube of the matrix $M_{gt}$,
is integer-valued, so we cannot exclude a possibility of 
a stone inflation for the golden tetrahedra with the
inflation factor $\tau^3$. 

An alternative proof
of the nonexistence of a stone inflation for the golden
tetrahedra is given in Subsection 3.3. It is based on the
analysis of irrationalities of areas of faces of
the golden tetrahedra. The analysis in Subsection 3.3
allows to show that a stone inflation for the golden tetrahedra
with the inflation factor $\tau^k$, $k=1,2,3,\dots$ cannot
exist for any $k$.

In Subsection 4.1 we present
the inflation rules for the {\em decorated} Mosseri--Sadoc tiles
(they are unions of the golden tetrahedra). These rules 
we obtain by a local derivation from the inflation
rules for the {\em decorated} golden tetrahedra 
(decoration increases
the number of tiles: there are eight decorated golden tetrahedra)
as the tiles of the projection class 
\ts2f,~\cite{PHK, HKP, PO}.

In Subsection 4.2 we show that the inflation matrix in the case
of the Mosseri--Sadoc tilings is uniquely reconstructed from the
volumes of the prototiles and their Dehn invariants.
Also, we explain in Subsection 4.2 that the inflation matrix
for the Mosseri--Sadoc tiles is induced by the inflation
matrix for the golden tetrahedra. 

For the calculation of the Dehn invariants of the golden
tetrahedra we use a Conway--Radin--Sadun theorem (Appendix).

\vskip .3cm
\section{Dehn invariants}

\vskip .6cm
The Dehn invariant of a polyhedron $P$ takes values in a ring
${\bf R}\otimes {\bf R}_\pi$ where ${\bf R}_\pi$ is the 
additive
group of residues of real numbers modulo $\pi$; the tensor
product is over ${\bf Z}$, the ring of rational integers.
Denote by $l_i$ the lengths of edges of $P$.
Denote by $\a_i$  the corresponding lateral angles and by
$\bar{\a_i}$ -- the residue classes of $\a_i$ modulo $\pi$.
The Dehn invariant, ${\cal D}(P)$, of the polyhedron $P$ is equal to
\be
{\cal D}(P)=\sum l_i\otimes \bar{\a_i}
\ ,
\lb{di}
\ee
with the sum over all edges of $P$.

Historically, Dehn invariants appeared in solving the
Hilbert's third problem~\cite{H} which asks whether one can
calculate the volume of a polyhedron without a limiting
procedure. More precisely, given two polyhedra of the same
volume, can one cut one and paste the pieces to build
another one? Or, is equality of volumes of two polyhedra
sufficient for their scissor equivalence?

Dehn~\cite{De} has shown that the quantity (\ref{di})
is scissor--invariant and gave an example of two polyhedra
of the same volume but having different Dehn invariants.
Thus, equality of Dehn invariants is a necessary condition
for the scissor equivalence. Later, Sydler~\cite{Sy} has shown
that in dimension 3 the equality of volumes and Dehn invariants
is also a sufficient condition for the scissor equivalence.
See~\cite{B} for more information on the Dehn invariants.

\vskip .3cm
\section{Inflation of golden tetrahedra}

\vskip .6cm
In this Section we discuss several aspects of the
inflation of the golden tetrahedra, not only the inflation
of these tiles as the prototiles {\em in} the projection
class of the tilings \ts2f.
The projection class of  the locally isomorphic tilings \ts2f \
and the inflation rules for the tiles in this class
have been considered in Refs.~\cite{KPZ,PHK,HKP}.

\vskip .5cm
\subsection{Golden tetrahedra and their Dehn invariants}

\vskip .5cm
%
%
\begin{figure}[h]
\centerline{
}
\caption{ (see Fig1.gif) 
The tiles of the projection class of the tilings
${\cal T}^{*(2F)}$: $G^*$, $F^*$, $ A^*$, $B^*$,
$C^*$ and $D^*$ (from left to right),
the golden tetrahedra.
All edges of the tetrahedra are parallel
to the 2fold symmetry axes of the icosahedron. They are of 
the standard
length \zfo \ (denoted by $1$ in the Figure) and $\tau$\zfo \
(denoted by $\tau$ in the Figure), 
\zfo \  $=\sqrt{\frac{2}{\tau + 2}}$.
The representative lateral angles are shown.
The ${\bf Z}_3$ 
rotational
symmetry of the tiles $G^*$ and $F^*$, the ${\bf Z}_2$
rotational symmetry of the tiles $A^*$ and $B^*$
and reflection symmetry of the tiles
$C^*$ and $D^*$ allow to reconstruct all other lateral angles.}
\end{figure}

Golden triangles are triangles with edge lengths $1$ and $\t$
(in some scale) satisfying the condition: not all edges
of a triangle are congruent.
There are two golden triangles: with edge lengths $(1,1,\t )$ and
with edge lengths $(1,\t ,\t )$.
A property of the golden triangles: edges of each of them
can be aligned in the plane parallelly to the symmetry 
axes of a given pentagon.

Golden tetrahedra are tetrahedra with edge lengths 1 and $\tau$
(therefore the faces of the golden tetrahedra can be either
golden or regular triangles) satisfying the condition:
not all faces of a tetrahedron are congruent. A property: 
golden tetrahedra are tetrahedra the edges of which can be 
aligned in the space parallelly to the 2fold symmetry axes 
of a given icosahedron. 

There could be seven golden tetrahedra but it turns
out that one of them is flat.
The six non--flat golden tetrahedra,
$G^*$, $F^*$, $A^*$, $B^*$, $C^*$ and $D^*$ are shown in Fig.~1.

All the lateral angles of the golden tetrahedra
are expressed in terms of four acute ($<\pi/2$) angles
$\alpha$, $\beta$, $\gamma$ and $\delta$,
\be \begin{array}{ccl}
\cos \alpha &=&\fracds{\t}{\t +2}=\fracds{1}{\sqrt{5}}\ ,\\[.4cm]
\cos \beta &=&\fracds{\t +1}{\sqrt{3}\sqrt{\t +2}}\ ,\\[.4cm]
\cos \gamma &=&\fracds{\t +2}{3\t}=\fracds{\sqrt{5}}{3}\ ,\\[.4cm]
\cos \delta &=&\fracds{\t -1}{\sqrt{3}\sqrt{\t +2}}\ .
\end{array}\lb{later}\ee
In ${\bf R}_\pi$ there are linear dependences between lateral
angles $\alpha$, $\beta$, $\gamma$ and $\delta$.

\vskip .5cm
\ni
{\bf Lemma 1.}
\ba
\a +\g +2\b &=&\pi\ ,\\[.2cm]
\a -\g +2\d &=&\pi\ .
\ea

\vskip .5cm
\ni
{\it Proof.} Straightforward. $\enb$

\vskip .5cm
Therefore, in ${\bf R}_\pi$ we have relations
\ba
\bar{\a}&=&-\bar{\b}-\bar{\d}\ ,\lb{dep1}\\[.2cm]
\bar{\g}&=&-\bar{\b}+\bar{\d}\ .\lb{dep2}
\ea
Next step is to prove that there are no more relations: 
in other
words, the images of angles $\b$ and $\d$ are independent in
${\bf R}_\pi$. Because of (\ref{dep1}) and (\ref{dep2}) it is
sufficient to check the
independence of $\bar{\a}$ and $\bar{\g}$.

\vskip .5cm
\ni
{\bf Lemma 2.} The images of angles
$\a$ and $\g$ in ${\bf R}_\pi$ are independent.

\vskip .5cm
\ni
{\it Proof.} For notation see Appendix.

The angles $\a$ and $\g$ are pure geodetic. One can check that
\be
\a =\langle 5\rangle_1 \ ,\ \g =
\frac{\pi}{2}-2\langle 3\rangle_5 \ .
\ee
The angles $\langle 5\rangle_1$ and $\langle 3\rangle_5$
are elements of the basis constructed by Conway--Radin--Sadun.
Thus, by the Conway--Radin--Sadun theorem (Appendix),
the angles $\a$ and $\g$ are independent. $\enb$

\vskip .5cm
The calculation of Dehn invariants of the golden tetrahedra is
now immediate. We shall use $\b$ and $\d$ as independent angles.
We express the Dehn invariants of the golden tetrahedra by the
vector $\vec{d}_{gt}$
\be
\vec{d}_{gt}={\cal D}
     \left(
\begin{array}{c}A^*\\B^*\\C^*\\D^*\\F^*\\G^*\end{array}\right)
     =\left( \begin{array}{c}
       -\t-1\\ \t+5\\3\t-2\\-2\t\\-3\t\\3\t+3\end{array}\right)
\otimes\bar{\b}+
      \left( \begin{array}{c}
       5\t-1\\ \t-1\\-2\\-2\t-3\\-3\t+3\\3\end{array}\right)
\otimes\bar{\d}
\ .
\lb{dgt}
\ee
The subscript $gt$ stands for ``golden tetrahedra".

The vector $\vec{v}_{gt}$ 
of volumes of the golden tetrahedra is
\be
\vec{v}_{gt}={\rm Vol}
     \left(
\begin{array}{c}A^*\\B^*\\C^*\\D^*\\F^*\\G^*\end{array}\right)
     =\frac{1}{12}\left( \begin{array}{c}
       2\t+1\\ 1\\ \t+1\\ \t\\ \t+1\\ \t\end{array}\right)
\ .
\lb{vgt}
\ee

\vskip .5cm
\subsection{On inflation of golden tetrahedra}

\vskip .5cm
First we show how to use the Dehn invariants as a necessary
condition for the existence of the stone inflation. By
definition, the inflation is ``stone"~\cite{D} if the 
inflated tiles
are composed of the whole original tiles; in other words,
one does not need to cut the original tiles into smaller pieces.
In particular, it follows that the volume matrix of the
stone inflation has integer entries.

\vskip .5cm
\ni
{\bf Lemma 1.} The golden tetrahedra as prototiles of a 
space tiling do not admit a stone inflation with an 
inflation factor
$\t$.

\vskip .5cm
\ni
{\it Proof.} Assume that the stone inflation exists.
Let $M_{gt}$ be its inflation matrix. Since the inflation 
is stone, the matrix elements of $M_{gt}$ are rational integers. 
In particular,
$M_{gt}$ is stable under the action of the
Galois group, $\t\rightarrow -1/\t$.

The vector $\vec{v}_{gt}$
(the vector of volumes of the tiles, eqn. (\ref{vgt}))
is an eigenvector of $M_{gt}$ with an eigenvalue $\t^3$.

The additivity of Dehn invariants implies that
the vector $\vec{d}_{gt}$
(the vector of Dehn invariants of the tiles, eqn. (\ref{dgt}))
is an eigenvector of $M_{gt}$ with an eigenvalue $\t$
(the eigenvalue is $\t$ because Dehn invariants have dimension
[length]$^1$).
Decomposing the vector of Dehn invariants in $\bar{\b}$ and
$\bar{\d}$ we obtain two eigenvectors of $M_{gt}$
with the eigenvalue $\t$.

Explicitely, we have for the volume vector:
\be
M_{gt}\left( \begin{array}{c}
    2\t+1\\1\\ \t+1\\ \t\\ \t+1\\ \t\end{array}\right) =
    \left( \begin{array}{c}
    8\t+5\\ 2\t+1\\ 5\t+3\\ 3\t+2\\ 5\t+3\\ 3\t+2\end{array}
    \right) 
\ ,
\lb{mvgt}
\ee
for the $\bar{\b}$--component of the Dehn vector:
\be
M_{gt}\left( \begin{array}{c}
    -\t-1\\ \t+5\\ 3\t-2\\ -2\t\\ -3\t\\ 3\t+3\end{array}
    \right) =
    \left( \begin{array}{c}
    -2\t-1\\ 6\t+1\\ \t+3\\ -2\t-2\\ -3\t-3\\ 6\t+3\end{array}
    \right) 
\ ,
\lb{mdgtb}
\ee
and for the $\bar{\d}$--component of the Dehn vector:
\be
M_{gt}\left( \begin{array}{c}
    5\t-1\\ \t-1\\ -2\\ -2\t-3\\ -3\t+3\\ 3\end{array}\right)
    =
    \left( \begin{array}{c}
    4\t+5\\ 1\\ -2\t\\ -5\t-2\\ -3\\ 3\t\end{array}\right) 
\ .
\lb{mdgtd}
\ee

The Galois automorphism $\t\rightarrow -1/\t$ produces three
more eigenvectors of $M_{gt}$. Since the entries of $M_{gt}$
are integer, to use the Galois automorphism is the same as
to decompose vector equalities (\ref{mvgt}), (\ref{mdgtb})
and (\ref{mdgtd}) in the powers of $\t$
({\it i.e.} consider $\t^0$--  and $\t^1$--components of
(\ref{mvgt}), (\ref{mdgtb}) and (\ref{mdgtd})).
Writing all the columns together we obtain a matrix equality,
\be
M_{gt}\left( \begin{array}{cccccc}
    2&1&-1&-1&5&-1\\
    0&1&1&5&1&-1\\
    1&1&3&-2&0&-2\\
    1&0&-2&0&-2&-3\\
    1&1&-3&0&-3&3\\
    1&0&3&3&0&3\end{array}\right) =
    \left( \begin{array}{cccccc}
    8&5&-2&-1&4&5\\
    2&1&6&1&0&1\\
    5&3&1&3&-2&0\\
    3&2&-2&-2&-5&-2\\
    5&3&-3&-3&0&-3\\
    3&2&6&3&3&0\end{array}\right)
\ .
\lb{eqi}
\ee
The matrix $M_{gt}$ is acting on a $6\times 6$ matrix whose 
first column is $\t^1$--component of (\ref{mvgt}),
the second column is $\t^0$--component of (\ref{mvgt});
the 3rd and 4th columns are $\t^1$-- and $\t^0$--components of
(\ref{mdgtb}); the 5th and 6th columns are $\t^1$--
and $\t^0$--components of (\ref{mdgtd}).

The eqn. (\ref{eqi}) is the matrix equation for the matrix
$M_{gt}$.
We found the complete basis of eigenvectors, therefore the
solution is unique and we find
\be
M_{gt}=\left( \begin{array}{cccccc}
    2&0&1&0&2&1\\
    0&0&1&0&0&1\\
    1/2&1/2&1&1&1&1\\
    0&0&1&1&1&0\\
    1&0&1&1&1&0\\
    1/2&1/2&1&0&0&1\end{array}\right) \ .
\lb{mgt}
\ee
The matrix entries of $M_{gt}$ are not integers therefore
a stone inflation with the inflation factor $\tau$
cannot exist. {\it Q. E. D.} \enb

\vskip .5cm
We actually proved more: we proved that if an inflation
with a {\em rational} inflation matrix existed then the inflation
matrix would necessarily be equal to (\ref{mgt}). In other 
words,
having assumed that the inflation matrix is rational we
could reconstruct it uniquely. This happened because of
a coincidence: $2\times 3=6$. Here 2 is the order of the
Galois group, 3 is the number of independent invariants
(the volume and the two Dehn invariants) while 6 is the 
number of
tiles. Due to this coincidence we obtained the matrix
equation for $M_{gt}$ admitting a unique solution. We
don't have a good explanation for this coincidence.

An inflation, with the inflation factor $\t$ for the
golden tetrahedra as the prototiles of the  projection class of
the  tilings \ts2f \  (obtained by the icosahedrally 
invariant projection from the $D_6$ 
lattice) 
has been found in Refs.~\cite{PHK,HKP}.
There, one has to divide the tiles $C^*$ and $G^*$, each
into two
subtypes: ``blue" and ``red", and these subtypes inflate
differently. Therefore, the number of tiles becomes 8.
The volume inflation matrix $M_{{\cal T}^{*(2F)}}$ is equal to
\be 
\left( \begin{array}{cccccccc}
    \! 11\t-16&\! 0&\! 2\t-2&\! 2\t-3&\! 0&\! 9\t-13&\! \t-1&\! 3\t-4\\
    \! 0&\! 0&\! 0&\! 1&\! 0&\! 0&\! 0&\! 1\\
    \! -2\t+4&\! 1&\! 0&\! -\t+2&\! 1&\! -\t+3&\! 0&\! -\t+2\\
    \! -9\t+15&\! 0&\! -2\t+4&\! -\t+2&\! 1&\! -8\t+14&\! -\t+2&\! -2\t+4\\
    \! 0&\! 0&\! 0&\! 1&\! 1&\! 1&\! 0&\! 0\\
    \! 1&\! 0&\! 1&\! 0&\! 1&\! 1&\! 0&\! 0\\
    \! -2\t+4&\! 1&\! 0&\! -\t+2&\! 0&\! -\t+2&\! 0&\! -\t+2\\
    \! -9\t+15&\! 0&\! -2\t+4&\! -\t+2&\! 0&\! -8\t+13&\! -\t+2&\! -2\t+4
\end{array}\right) 
\lb{m2f}
\ee
in the following ordering of the tiles: $ A^*$, $B^*$,
$C^{*b}$, $C^{*r}$, $D^*$, $F^*$, $G^{*b}$
and $G^{*r}$.  The upper indices ``$b$" and ``$r$" denote
the ``blue" and the ``red" variants of tiles, respectively.

It is interesting to note that: 1. for the tiles $B^*$, $D^*$ and
$F^*$ the inflation matrices $M_{gt}$ and 
$M_{{\cal T}^{*(2F)}}$
give the same results (up to colors); 2. noninteger entries in
(\ref{mgt}) appear exactly in the columns corresponding to the 
tiles $C^*$ and $G^*$ -- the tiles which are getting blue and red 
colors in the inflation with the matrix (\ref{m2f}).

Lemma 1 of this Subsection shows that a stone inflation 
with the inflation factor $\tau$ is impossible. We could however
try to construct a hypothetic inflation matrix with
an inflation factor $\tau^k$ with integer positive $k$, $k>1$.
As in the proof of the Lemma 1, the volume vectors and the vectors of
Dehn invariants fix the inflation matrix uniquely: the only
possible inflation matrix with the inflation factor $\tau^k$
can be the matrix $M_{gt}^k$. It turns out that there are powers
of the matrix $M_{gt}$ which are integer-valued.

\vskip .5cm
\ni {\bf Lemma 2.} The matrix $M_{gt}^k$ has integer entries
if and only if $k$ is divisible by 3.

\vskip .5cm
\ni {\it Proof.} A direct calculation gives
\be
M_{gt}^2=\left( \begin{array}{cccccc}
    7&1&6&3&7&4\\
    1&1&2&2&1&2\\
    3&1&5&3&4&3\\
    3/2&1/2&3&3&3&1\\
    7/2&1/2&4&3&5&2\\
    2&1&3&1&2&3\end{array}\right) 
\ee
and
\be
M_{gt}^3=\left( \begin{array}{cccccc}
    26&5&28&16&30&18\\
    5&2&8&4&6&6\\
    14&4&19&12&18&12\\
    8&2&12&9&12&6\\
    15&3&18&12&19&10\\
    9&3&12&6&10&9\end{array}\right) \ .
\ee
Thus, $M_{gt}^3$ is an integer-valued matrix and therefore
matrices $M_{gt}^{3k}$ are integer-valued as well.

It is left to prove that if $n$ is not a multiple of 3
then $M_{gt}^n$ is not integer-valued.

By construction, the eigenvalues of $M_{gt}$ are 
\be \tau^3\ ,\ (-\tau^{-3})\ ,\ \tau\ {\mathrm{and}}\  (-\tau^{-1})
\ .\ee
Therefore, the minimal polynomial for $M_{gt}$ is
\be \chi (x)=x^4-5x^3+2x^2+5x+1\ ,\ee
$\chi (M_{gt})=0$.

A straightforward check shows that if 
\be x^4=5x^3-2x^2-5x-1 \ee
then
\be x^n=a_nx^3+b_nx^2+c_nx+d_n\ee
with
\be a_n=\frac{1}{3}\left( \frac{f_{3(n-1)}}{2}-f_{n-1} 
   \right)\ee
and
\be \begin{array}{ccl}
  b_n&=&a_{n+1}-a_n\ ,\\[.2cm]
  c_n&=&-a_{n+1}+3a_n+f_n\ ,\\[.2cm]
  d_n&=&-a_{n+1}+4a_n+f_{n-1}\ .
\end{array}\ee
Here $\{ f_n \}$ are Fibonacci numbers defined by: $f_0=0$, $f_1=1$ and 
$f_{n+1}=f_n+f_{n-1}$.

Therefore,
\be M_{gt}^n=a_nM_{gt}^3+b_nM_{gt}^2+c_nM_{gt}+d_n\, {\mathrm{\bf Id}}
\ ,\ee
where ${\mathrm{\bf Id}}$ is the unit matrix.

The numbers $a_n$, $b_n$, $c_n$ and $d_n$ are integer. The matrices
$M_{gt}^3$ and ${\mathrm{\bf Id}}$ have integer entries. The
matrices $M_{gt}$ and $M_{gt}^2$ have -- at different places --
rational entries with the denominator 2. Therefore, the matrix
$M_{gt}^n$ has integer entries if and only if the integers $b_n$ and
$c_n$ are even which means that
\be a_{n+1}\equiv a_n\ ({\mathrm{mod}}\ 2) \lb{em1}\ee
and
\be -a_{n+1}+3a_n+f_n\equiv 0\ ({\mathrm{mod}}\ 2) \ .\lb{em2}\ee
Substitution of (\ref{em1}) into (\ref{em2}) gives 
$f_n\equiv 0\ ({\mathrm{mod}}\ 2)$. It is well known that $f_n$
is even if and only if $n$ is a multiple of 3 (see, e.g.,
\cite{GKP}, Chapter 6). $\enb$ 

\vskip .5cm
To conclude: with the help of the Dehn invariants one is
able to show that a stone inflation with the inflation factor
$\tau$ is impossible. However one cannot exclude a stone inflation
with the inflation factor $\tau^3$.

\vskip .5cm
In the next Subsection we shall show, using a different method,
that a stone inflation with the inflation factor $\tau^3$ is 
impossible as well.

\vskip .8cm
\subsection{Faces of golden tetrahedra}

\vskip .5cm
The Lemma 1 proved in Subsection 3.2 
shows that the stone inflation with the inflation factor $\tau$
is impossible due 
to the scissor invariants of the tiles -- the volumes and
the Dehn invariants. 

Here we shall give another argument showing the impossibility
of a stone inflation. This argument uses the geometry of
faces of the tiles.

More precisely, using Dehn invariants amounts to analyzing
irrationalities in the lateral angles of the golden tetrahedra.
Now we shall analyze irrationalities in the areas of the faces of
the golden tetrahedra.

The faces of the golden tetrahedra are golden and regular
triangles.

Denote the regular triangle, with the edge length 1, by $\D_r$,
the acute golden triangle (with edge lengths $\tau$, $\tau$ and 1)
by $\D_a$ and the obtuse golden triangle (with edge lengths
$\tau$, 1 and 1) by $\D_o$.

For an arbitrary triangle $\D$, a notation $\tau^{-k}\D$ means
the triangle $\D$ scaled by $\tau^{-k}$. Also, for a triangle
$\D$, denote a set of triangles $\{ \tau^{-k}\D ,\ k=1,2,3,\dots\} $
by $\tau^-\D$.

The areas $A(\D )$ of the triangles are
\be \begin{array}{ccl}
  A_r&\equiv& A(\D_r)=\fracds{\sqrt{3}}{4}\ ,\\[.3cm]
  A_o&\equiv& A(\D_o)=\fracds{\sqrt{\tau +2}}{4}\ ,\\[.3cm]
  A_a&\equiv& A(\D_a)=\fracds{\tau \sqrt{\tau +2}}{4}\ .
\end{array}\lb{areas}\ee
The $\sqrt{\ \ }$ will always denote the positive branch of
the square root.

It is interesting to note that the irrationalities in the
areas of the faces are exactly the same as in trigonometric functions
of the lateral angles (see (\ref{later})): $\sqrt{3}$, $\tau$
and $\sqrt{\tau +2}$.

We first prove an intuitively obvious technical Lemma
which shows that irrationalities expressing the areas 
(\ref{areas}) are different.

\vskip .5cm
\ni {\bf Lemma 1.} The irrationalities $\sqrt{3}$ and 
$\sqrt{\tau +2}$ are independent over the field $\ku [\tau ]$.

\vskip .5cm
\ni {\it Proof.} The number $\rho =\sqrt{\tau +2}$ satisfies
an equation $f(\rho )=0$ where
\be f(x)=x^4-5x^2+5\ .\ee
By the Eisenstein criterion (see, e.g., \cite{Cl}, Chapter 3),
the polynomial $f$ is irreducible over $\ku$. Moreover, $f$ splits
in $\ku [\rho ]$: its roots are
\be \pm\sqrt{\tau +2}\ \ {\mathrm{and}}\ \ \pm\sqrt{3-\tau}\ .\ee
The irrationality $\sqrt{3-\tau}$ belongs to the field $\ku [\rho ]$:
one has
\be \sqrt{3-\tau} =\tau^{-1}\sqrt{\tau +2}\in \ku [\rho ]\ .\ee
A splitting field of any polynomial is a Galois extension
(\cite{Cl}, Chapter 4). Therefore, the field $\ku [\rho ]$ --
as the splitting field of the polynomial $f$ -- is the Galois
extension of $\ku$. 

The automorphism group $Gal(\ku [\rho ]/\ku )$ is isomorphic to
the cyclic group ${\bf Z}_4$, with the generator $\s$,
\be \s :\sqrt{\tau +2}\rightarrow\sqrt{3-\tau}\ .\ee
In a basis 1, $\sqrt{\tau +2}$, $\tau$ and $\sqrt{3-\tau }$
of $\ku [\rho ]$ over $\ku$, the action of $\s$ on the other
elements of the basis is given by
\be \s :\tau\rightarrow -\tau^{-1}\ \ {\mathrm{and}}\ \ 
    \s:\sqrt{3-\tau }\rightarrow -\sqrt{\tau +2}\ .\ee
Hence, $\s^4=1$.

By the Fundamental Theorem of Galois Theory (see, e.g., \cite{Cl},
Chapter 4), a quadratic extension of $\ku$ between $\ku$ and
$\ku [\rho ]$ can be only the fixed field of $\s^2$ which is 
$\ku [\tau ]$.

In particular, $\sqrt{3}\not\in \ku [\rho]$ 
(since, clearly, $\sqrt{3}\not\in \ku [\tau ]$). $\enb$ 

\vskip .5cm
{\it Remark.} It is also easy to prove in an elementary way that 
the equation $x^2=3$ does not have solutions in $\ku [\rho ]$.

\vskip .5cm
\ni {\bf Corollary.} The field $\ku [\rho ,\sqrt{3}]$ admits an
automorphism $\phi$ which satisfies: 

1. $\phi :\sqrt{3}\rightarrow -\sqrt{3}$; 

2. the fixed field of $\phi$ is $\ku [\rho ]$. 

\vskip .5cm
\ni {\it Proof.} It follows from the Lemma 1 that the field 
$\ku [\rho ,\sqrt{3}]$ is a quadratic extension of the field 
$\ku [\rho ]$. In characteristic 0, any quadratic extension is
Galois (\cite{Cl}, Chapter 4). This immediately implies the existence 
of the automorphism $\phi$. $\enb$ 

\vskip .5cm
We shall now apply these algebraic preliminaries to the analysis
of a stone inflation.

If a stone inflation existed, the faces of inflated tiles
would be covered by the faces of the original tiles.

\vskip .5cm
\ni {\bf Lemma 2.} 1. Assume that a regular triangle $\D_r$ is
covered by a finite (interior)-disjoint union of regular triangles
from $\tau^-\D_r$ and golden triangles from $\tau^-\D_a$
and $\tau^-\D_o$. Then the golden triangles are absent in the
covering. In other words, a regular triangle can be covered 
by regular triangles only.

\vskip .2cm
2. Similarly, the golden triangles can be covered by the golden
triangles only, the regular triangles must be absent in
the covering.

\vskip .5cm
\ni {\it Proof.} Suppose that the triangle $\D_r$ is covered by a finite
union of triangles from $\tau^-\D_r$, $\tau^-\D_a$ and $\tau^-\D_o$.
Then for the areas we have
\be A_r=p_1(\tau^{-2})A_r+p_2(\tau^{-2})A_a+p_3(\tau^{-2})A_o\ ,
\lb{eqqq}\ee
where $p_1$, $p_2$ and $p_3$ are polynomials with nonnegative
integer coefficients and the polynomial
$p_1$ does not have a constant term.

Let $X=\sqrt{3} (1-p_1(\tau^{-2}))$ and 
$Y=p_2(\tau^{-2})\tau\sqrt{\tau +2}+p_3(\tau^{-2})\sqrt{\tau +2}$.
The equality (\ref{eqqq}) is equivalent to $X=Y$.

Applying the automorphism $\phi$ (Corollary, Lemma 1) to the equality
$X=Y$ we find $(-X)=Y$ and it follows that $X=0$ and $Y=0$ separately. 
Since each term in the expressions $p_2(\tau^{-2})A_a$ and
$p_3(\tau^{-2})A_o$ is nonnegative, the equality $Y=0$ implies
that the polynomials $p_2(x)$ and $p_3(x)$ are identically
zero. This means that the golden triangles 
are absent. 

The considerations with coverings of the golden triangles are analogous.
$\enb$

\vskip .5cm
To prove the nonexistence of a stone inflation we shall consider
coverings of the regular triangle.

We shall prove that a regular triangle with the edge
of length $\t^k$ cannot be covered by regular triangles
with the edge lengths $\t^i$, $i=0,\dots ,k-1$. 
This will imply that there is no stone inflation with the 
inflation factor $\t^k$ for any $k$.

In fact, the same arguments can be applied to coverings
of any triangle $\D$ by $\tau^k$--smaller copies of the same 
triangle.

Consider an arbitrary triangle $\D$. Suppose that the
triangle $\tau^k\D$
is divided into a finite (interior)--disjoint union of 
triangles $\tau^i\D$ with $i=0,\dots ,k-1$.
Consider such division with a smallest possible $k$. Then a
triangle $\D =\t^0\D$ is necessarily
present -- otherwise,
rescaling by $1/\t$ we would obtain the division of the 
triangle $\t^{k-1}\D$ contradicting to the minimality of $k$.

Denote by $\a_i$ the number of triangles $\t^i\D$.
We have $\a_i\geq 0$ for $i=1,\dots ,k-1$ and $\a_0>0$. Put
$\s=\t^2$. From the area consideration it follows that
\be 
\s^k =\a_{k-1}\s^{k-1}+\dots +\a_0\ .
\ee

It is this statement which will lead to a contradiction.

\vskip .5cm
\ni
{\bf Lemma 3.} The number $\s$ cannot satisfy an equation
\be 
\s^k -\a_{k-1}\s^{k-1}-\dots -\a_0 =0\ ,
\ee
where $\a_1,\dots ,\a_{k-1}$ are nonnegative integer 
numbers and
$\a_0$ is a positive integer number.

\vskip .5cm
\ni
{\it Proof.}
The minimal equation (over ${\bf Z}$) for
$\s=\frac{3+\sqrt{5}}{2}$ is 
\be \s^2-3\s+1=0\ .\ee

Let $p(x)=x^k-\a_{k-1}x^{k-1}-\dots -\a_0$.
Assume that $p(\s )=0$.
This means that one can divide $p(x)$ by $x^2-3x+1$:
\be 
p(x)=(x^{k-2}+\b_{k-3}x^{k-3}+\dots +\b_0)(x^2-3x+1)\ .
\ee
Collecting coefficients in powers of $x$ we obtain the 
following system:
\be
\begin{array}{lcl}
    -\a_{k-1}&=&-3+\b_{k-3}\\[.2cm]
    -\a_{k-2}&=&1-3\b_{k-3}+\b_{k-4}\\[.2cm]
    -\a_{k-3}&=&\b_{k-3}-3\b_{k-4}+\b_{k-5}\\[.2cm]
             &\vdots& \\
    -\a_2&=&\b_2-3\b_1+\b_0\\[.2cm]
    -\a_1&=&\b_1-3\b_0\\[.2cm]
    -\a_0&=&\b_0 \end{array}
\lb{rec}
\ee
Let $\psi_n=f_{2n+2}$ where $f_n$ are Fibonacci numbers. 
Then we have
$\psi_0=1$, $\psi_1=3$ and
\be
\psi_{n+1}=3\psi_n-\psi_{n-1}
\ .
\lb{recpsi}
\ee

Let $S=\a_{k-1}\psi_0+\a_{k-2}\psi_1+\dots +\a_0\psi_{k-1}$.

Substituting expressions for $\a_i$ from (\ref{rec}) one finds
that due to (\ref{recpsi}) the terms with $\psi_i$ for $i>1$
cancel and one is left with
\be
S=-3\psi_0+\psi_1\equiv -3+3=0\ ,
\ee
which is impossible since all $\psi_i$ are positive, $\a_i$
are nonnegative and $\a_0$ is positive. $\enb$

\vskip .5cm
As we have seen, Lemma 3 implies the following statement.

\vskip .5cm
\ni {\bf Corollary.} A regular triangle cannot be covered by 
$\tau^k$--smaller regular triangles.

\vskip .5cm
With these preliminaries we are now prepared to show
that a stone inflation for the golden tetrahedra is impossible.

\vskip .5cm
\ni {\bf Proposition.} For the golden tetrahedra, a stone 
inflation with the inflation factor $\tau^k$, with an arbitrary
positive integer $k$, does not exist.

\vskip .5cm
\ni {\it Proof.} As it was said above, an existence of a stone 
inflation implies that the faces of the inflated tiles can be 
covered by the faces of the tiles of the original size.

In particular, a face which is an inflated regular triangle, would be
covered by regular and golden triangles.

Lemma 2 shows that the golden triangles cannot appear in such
covering. Therefore, the inflated regular triangle can be covered
by regular triangles only -- which is impossible by Corollary,
Lemma 3.

This contradiction shows that a stone inflation does not exist.
$\enb$

\vskip .5cm
{\it Remark.} The known tilings $\tst2f$ have the following property.
The golden tetrahedra in the tiling of the space have their edges
parallel to the 2fold symmetry axes of the icosahedron (``the
long range orientational order'').
The faces of the tiles which are regular triangles are all located in
the planes perpendicular to the 3fold symmetry axes of
the icosahedron. However the golden triangles are all
perpendicular to the 5fold symmetry axes. Therefore if a stone inflation 
for the tilings $\tst2f$ existed, the regular triangles could be 
covered only by the smaller regular triangles due to the
orientation of the faces. In this case we don't need
Lemmas 1 and 2.

\vskip .5cm
We stress again that an existence of the ``rational"
inflation rules for the golden tetrahedra (eqn. \ref{mgt})
is hypothetic because in our algebraic approach we do not
impose any restriction on the orientations of the tiles in 
the tiling of the 3dimensional space.

\vskip .5cm
The logic used in this Subsection gives an additional
motivation to consider minimal packages of the golden tetrahedra
in which the regular faces are all hidden (see Section 4).

\vskip .3cm
\section{Mosseri--Sadoc tiles}

\vskip .6cm
%
%
\begin{figure}[h]
\centerline{
\hfill
\hfill
}
\vfill
\centerline{
\hfill
\hfill
}
\caption{ (see Fig2.gif) 
The outer shape of the ``window" $W (= V_\perp)$ 
of the projection class of the tilings ${\cal T}^{(MS)}$ in \es.
It is the triacontahedron with an edge length \ffo$=1/\sqrt{2}$, 
the standard length parallel to the 5fold symmetry axes of the  
icosahedron. 
The Figure shows the tiles $a$, $r$, $m$, $s$ and $z$ in \ep. 
The symmetries 
of the tiles and the representative lengths of edges 
are marked. In this paper the standard length 
\zfo \ $=\sqrt{\frac{2}{\tau + 2}}$ is set to 1.}
\end{figure}
%
%
\begin{figure}[h]
\centerline{
}
\caption{ (see Fig3.gif) 
The tiles $r$ and $m$ appear in  the projection
class  of the tilings \tms \ always
together as the union $h$, $h = r\bigcup m$. }
\end{figure}
%
%
\begin{figure}[h]
\centerline{
\hfill
}
\vfill
\centerline{
\hfill
\hfill
}
\caption{ (see Fig4.gif) 
The tiles  $a$, $m$, $r$, $z$ and $s$ are obtained 
by the packing of the golden tetrahedra.
}
\end{figure}

The five prototiles, $a$, $m$, $r$, $z$ and $s$ of the
projection class of  the tilings \tms (see \cite{KP}) are shown 
in Fig.~2.
The tiles $r$ and $m$ appear in \tms \ always together as
a tile $h$, $h = r \bigcup m$, see Fig.~3.
The prototiles $z$, $h$, $s$ and $a$ are of the same shape as
the prototiles of the inflation class of the tilings 
introduced by
Sadoc and Mosseri~\cite{MS}, and we call them the Mosseri--Sadoc
tiles. The tiles $a$, $m$, $r$, $z$ and $s$
are composed  of the golden
tetrahedra~\cite{MS,KP}, as shown in Fig.~4, in such a way that
the  regular triangles of the golden tetrahedra
are all hidden~\cite{KP}. Hence, the faces of the composed
tiles $a$, $m$, $r$, $z$ and $s$ are golden triangles only.
The same is true for the Mosseri--Sadoc tiles
$z$, $h$, $s$ and $a$.

Using additivity of Dehn invariants one finds the vector
of Dehn invariants for the Mosseri--Sadoc tiles:
\be
\vec{d}_{MS}={\cal D}
     \left( \begin{array}{c}z\\h\\s\\a\end{array}\right)
     =-5 \left( \begin{array}{c}
       \t\\ 2\\ \t-1\\-\t\end{array}\right)\otimes\bar{\a}\ .
\lb{dms}
\ee
Thus, the space of Dehn invariants for the Mosseri--Sadoc tiles
becomes 1dimensional, only the combination
$\bar{\a}=-\bar{\b}-\bar{\d}$ appears.

For the vector of volumes for the Mosseri--Sadoc tiles one
obtains
\be
\vec{v}_{MS}={\rm Vol}
     \left( \begin{array}{c}z\\h\\s\\a\end{array}\right)
     =\frac{1}{12}\left( \begin{array}{c}
       4\t+2\\ 6\t+4\\ 4\t+3\\ 2\t+1\end{array}\right)
\ .
\lb{vms}
\ee

{\it Note.} The Mosseri--Sadoc tile $h$ is the union of the
tiles $m$ and $r$ introduced in Ref.~\cite{KP}.
The volumes and the Dehn invariants of the tiles $m$ and $r$
are
\be
{\rm Vol}(m)=\frac{1}{12}(2\t+3)\ ,\
    {\rm Vol}(r)=\frac{1}{12}(4\t+1)\ .
\ee
\vskip .2cm
\be
{\cal D}(m)=5(\t-1)\otimes\bar{\a}\ ,\
    {\cal D}(r)=-5(\t+1)\otimes\bar{\a}\ .
\lb{demr}\ee
The Dehn invariants of both of them contain only the 
combination
$\bar{\a}$. Thus, were the tiles $m$ and $r$ not always glued
together, we wouldn't be able to write a matrix equation for
the inflation of 5 tiles $z$, $m$, $r$, $s$ and $a$.
That the tiles $m$ and $r$ in the projection class of the tilings
\tms \ do appear always together as the prototile $h$
has been shown in Ref.~\cite{KP} by the arguments of the 
projection method expressed in the ``orthogonal space". 
For the overview of the space tilings obtained by the projection
method see Ref.~\cite{PK}.

\vskip .5cm
\subsection{Inflation of decorated Mosseri--Sadoc tiles}

\vskip .5cm
%
%
\begin{figure}[h]
\centerline{
}
\caption{ (see Fig5.gif) 
The inflation rule for the decorated tile $a$:
$\tau a = a \bigcup s \bigcup a$. The ``white" arrow
marks the edge $\tau^2$\zfo, the ``long" edge
in the $\tau$\ts2f--class of the tilings (the \ts2f--class
of the tilings scaled by $\tau$).}
\end{figure}
%
%
\begin{figure}[h]
\centerline{
}
\caption{ (see Fig6.gif) 
The inflation rule for the decorated tile $m$:
$\tau m = a \bigcup s \bigcup z \bigcup a $.
The white arrow is marking the ``long" edge
in the $\tau$\ts2f--class of tilings. }
\end{figure}
%
%
\begin{figure}[h]
\centerline{
}
\caption{ (see Fig7.gif) 
The inflation rule for the decorated tile $r$:
$\tau r = z \bigcup s \bigcup m \bigcup r$. }
\end{figure}

Mosseri and Sadoc have given the inflation rules for their
$z$, $h$, $s$ and $a$ tiles~\cite{MS}. These rules were for the
stone inflation~\cite{D} of the tiles.
The inflation factor is
$\t=\frac{1+\sqrt{5}}{2}$.
The inflation matrix of the stone inflation of the tiles is
the matrix with integer coefficients.
It has been given by Sadoc and Mosseri~\cite{MS}
\be
M=\left( \begin{array}{cccc}
    1&1&1&1\\
    2&1&2&2\\
    1&1&1&2\\
    0&0&1&2
\end{array} \right)
\ ,
\lb{mmosa}
\ee
in the following ordering of the tiles: $z$, $h$, $s$ and $a$.

In the case of the Mosseri--Sadoc tiles, the stone inflation
is breaking the symmetry of the tiles.
The authors of~\cite{MS} haven't given a decoration of the
tiles which would  take care about the symmetry breaking and
{\em uniquely} define the inflation--deflation procedure
at {\em every step}.
In~\cite{KP} it has been shown that the projection class
of the locally isomorphic tilings \ts2f (see \cite{KPZ})
can be locally transformed
into the tilings \tms, \ts2f $\longrightarrow $ \tms.
The class \tms \ of the locally isomorphic tilings of the space 
by the Mosseri--Sadoc tiles has been defined by the icosahedral 
projection from the  $D_6$--lattice~\cite{KP}.
The important property is that {\em minimal} packages of
the six golden tetrahedra in \ts2f, satisfying the condition
that their equilateral faces (orthogonal to the 3fold
directions) are covered, lead to five tiles
$a$, $s$, $z$, $r$ and $m$~\cite{KP}.
Moreover, the tiles $r$ and $m$ appear always as the union
$r\bigcup m$, that is, the tile $h$ of Sadoc and Mosseri with
three mirror symmetries~\cite{KP}. See Figs.~2, 3 and 4.

%
%
\begin{figure}[h]
\centerline{
}
\caption{ (see Fig8.gif) 
The inflation rule for the decorated tile $z$:
$\tau z = \tau r \bigcup a $. The white arrows are marking the
``short" and the ``long" edges
in the  $\tau$\ts2f--class of tilings.}
\end{figure}
%
%
\begin{figure}[h]
\centerline{
}
\caption{ (see Fig9.gif) 
The inflation rule for the decorated tile $s$:
$\tau s = \tau z \bigcup a $. The white arrow is marking the
``long" edge
in the  $\tau$\ts2f--class of tilings. }
\end{figure}

It is apriori not evident that the inflation rules for the
Mosseri--Sadoc tiles in the projection class of the tilings
\tms \  are  the {\em same} as those suggested by Sadoc and
Mosseri~\cite{MS}.

The inflation rules for the \ts2f--tiles  {\em in} the projection
class
of the tilings \ts2f \ have been obtained  in 
Refs.~\cite{PHK,HKP}.
The inflation rules for
the prototiles in a projection class of tilings
are determined in the orthogonal space by a
procedure explained in Refs.~\cite{BKSZ,PHK}.
All edges of the \ts2f--tiles are
carrying the arrows and some of these arrows are
uniquely defining the inflation rules for
the \ts2f--tiles~\cite{PHK}.
By a local derivation of \tms \  from \ts2f, the Mosseri--Sadoc
tiles inherit these arrows~\cite{PO}.
The arrows which break the symmetry of
\tms--tiles are defining the inflation procedure uniquely.
The inflation--deflation rules for the decorated
$a$, $m$, $r$, $z$ and $s$ tiles {\em in} the projection class
of the tilings \tms \ are obtained through the local derivation
from the inflation--deflation rules for the {\em decorated}
golden tetrahedra (eight prototiles!)
as the tiles of  the projection class \ts2f.
We give the inflation rules for $a$, $m$, $r$, $z$ and $s$ 
tiles in Figs.~5 to 9.
If we keep in mind that the tiles $m$  and $r$
appear in \tms \ together as $h$, $m\bigcup r=h$, these are the
inflation--deflation rules for the projection class of
the tilings \tms \ of the space by the decorated
Mosseri--Sadoc tiles $z$, $h$, $s$ and $a$.
We see that the inflation rules for \tms \ as a {\em projection
specie}~\cite{KP} are the {\em same} (up to the decoration) 
as for the {\em inflation specie} given by Mosseri and 
Sadoc~\cite{MS}. 
By the fact that only the {\em decorated} 
Mosseri--Sadoc tiles {\em do} have the uniquely defined 
inflation--deflation procedure and by the fact that 
the inflation rules for the projection and inflation 
species  are the same,
we identify  the inflation~\cite{MS} and the
projection species~\cite{KP} and denote them  by
the same symbol, \tms.

\vskip .5cm
\subsection{Dehn invariants and stone inflation of
Mosseri--Sadoc tiles}

\vskip .5cm
In this Section we show that the inflation matrix for
the Mosseri--Sadoc tiles, $z$, $h$, $s$ and $a$,
can be uniquely reconstructed
from the Dehn invariants (and the volume).

Denote the inflation matrix by $M_{MS}$.

The vectors $\vec{d}_{MS}$ and $\vec{v}_{MS}$
(see eqns. (\ref{dms}) and (\ref{vms})) are eigenvectors of
the inflation matrix, with the eigenvalues $\t$ and $\t^3$
correspondingly (we remind that the eigenvalue is equal
to the inflation factor to the power which is the dimension
of the corresponding invariant).

Explicitely, for the vector of volumes  we have
\be
M_{MS}\left(
    \begin{array}{c}4\t+2\\6\t+4\\4\t+3\\2\t+1\end{array}\right)
   =\left(
    \begin{array}{c}16\t+10\\26\t+16\\18\t+11\\8\t+5
    \end{array}\right)
\lb{eigv}
\ee
and for the the vector of Dehn invariants:
\be
M_{MS}\left(\begin{array}{c}\t\\2\\ \t-1\\-\t
\end{array}\right)
   =\left(\begin{array}{c}\t+1\\2\t\\1\\-\t-1
\end{array}\right) \ .
\lb{eigd}
\ee

As for the golden tetrahedra tiles, assume that the inflation
matrix is rational. Then, applying the Galois automorphism
one finds two more eigenvectors of $M_{MS}$.
Again, as for tetrahedra, this amounts to the decomposition
of (\ref{eigv}) and (\ref{eigd}) in powers of $\t$.

Together, the four vector equations imply a matrix equation
\be
M_{MS}\left(\begin{array}{cccc}
     4&2&1&0\\
     6&4&0&2\\
     4&3&1&-1\\
     2&1&-1&0\end{array}\right)
   =\left(\begin{array}{cccc}
     16&10&1&1\\
     26&16&2&0\\
     18&11&0&1\\
     8&5&-1&-1\end{array}\right)
\ .
\lb{mms}
\ee
The solution of this equation is unique and we rediscover the
matrix (\ref{mmosa}).

Note that as for the tetrahedra, the uniqueness happens
because of the coincidence: the number of tiles equals
to the number of invariants times the order of the Galois
group.

\vskip .5cm
{\it Remarks.} 1. The inflation matrix $M_{MS}$ for the
Mosseri--Sadoc tiles is ``induced'' by the inflation matrix $M_{gt}$
for the golden tetrahedra in the following sense.

Denote by $V_{gt}$ a six-dimensional vector space
with a basis
\be \{ e_{A^*},e_{B^*},e_{C^*},e_{D^*},e_{F^*},e_{G^*} \} \ee
labeled by the golden tetrahedra. The matrix $M_{gt}$ acts
in the vector space $V_{gt}$ in an obvious way. We shall denote
the corresponding operator by the same symbol $M_{gt}$.
The lattice $L_{gt}$ generated by the basis vectors is not preserved by 
the operator $M_{gt}$ since the entries of $M_{gt}$ are not integers.

Denote by $V_{MS}$ a four-dimensional vector space
with a basis
\be \{ e_{z},e_{h},e_{s},e_{a} \} \lb{basms}\ee
labeled by the Mosseri--Sadoc tiles. The basis vectors
generate a lattice $L_{MS}$.

A map $\psi_{gt}: V_{MS}\rightarrow V_{gt}$ given by
\be \begin{array}{ccl}
\psi_{gt}(e_z)&=&e_{A^*}+e_{C^*}+e_{G^*}\ ,\\[.2cm]
\psi_{gt}(e_h)&=&e_{A^*}+e_{B^*}+2e_{F^*}+2e_{G^*}\ ,\\[.2cm]
\psi_{gt}(e_s)&=&e_{A^*}+2e_{C^*}\ ,\\[.2cm]
\psi_{gt}(e_a)&=&e_{D^*}+e_{F^*}
\end{array}\lb{psigt}\ee
is an embedding. It is compatible with the lattice structure.

The map $\psi_{gt}$ reflects the way of packing the golden
tetrahedra into the Mosseri--Sadoc tiles (see Fig. 4).

A direct inspection shows that the four-dimensional subspace
Im$(\psi_{gt})$ of $V_{gt}$ is invariant under the action of 
$M_{gt}$ and the matrix of the induced operator in $V_{MS}$,
written in the basis (\ref{basms}), coincides with $M_{MS}$.

This is quite natural since both matrices, $M_{gt}$ and $M_{MS}$,
are uniquely determined by the geometrical data --
the volumes and the Dehn invariants.

\vskip .5cm
2. The space $V_{MS}$ is a subspace in a five-dimensional
space $V_{MS}'$ with a basis
\be \{ e_{z},e_{m},e_r,e_{s},e_{a} \} \ .\ee
The element $e_h$ is expressed as $e_h=e_m+e_r$.

The space $V_{MS}'$ also maps into $V_{gt}$, the second line
in (\ref{psigt}) gets replaced by
\be \begin{array}{ccl}
\psi_{gt}(e_m)&=&e_{B^*}+2e_{F^*}\ ,\\[.2cm]
\psi_{gt}(e_r)&=&e_{A^*}+2e_{G^*}\ . 
\end{array}\ee
It is not an embedding any more:
\be \psi_{gt}(e_r+e_s)=\psi_{gt}(2e_z)\ .\ee
This explains again (see eqs. (\ref{demr}) and the comment after
them) that the inflation matrix for the
five tiles $a$, $m$, $r$, $z$ and $s$ cannot be reconstructed
from the Dehn invariants and the volumes (in other words, from 
the matrix $M_{gt}$).

In fact, the inflation matrix for the tiles $a$, $m$, $r$, $z$ and 
$s$ which reads (in this ordering of the tiles)
\be \left( \begin{array}{ccccc}
  2&0&0&0&1\\
  2&0&0&1&1\\
  0&1&1&1&1\\
  1&1&1&1&1\\
  2&1&1&1&1
\end{array} \right) \ee
is degenerate, so it cannot be induced by the nondegenerate
matrix $M_{gt}$.
  
\vskip .5cm
3. Denote by $V_{\tst2f}$ an eight-dimensional vector space
with a basis
\be \{ \ti{e}_{A^*},\ti{e}_{B^*},\ti{e}_{C^{*b}},\ti{e}_{C^{*r}},
  \ti{e}_{D^*},\ti{e}_{F^*},\ti{e}_{G^{*b}},\ti{e}_{G^{*r}} \} \ee
labeled by the coloured golden tetrahedra. The matrix $M_{\tst2f}$ 
becomes an operator acting in the space $V_{\tst2f}$.

Define a map $\psi_{\tst2f}: V_{MS}\rightarrow V_{\tst2f}$ by
\be \begin{array}{ccl}
\psi_{\tst2f}(e_z)&=&\ti{e}_{A^*}+\ti{e}_{C^{*b}}+\ti{e}_{G^{*r}}\ ,\\[.2cm]
\psi_{\tst2f}(e_h)&=&\ti{e}_{A^*}+\ti{e}_{B^*}+
   2\ti{e}_{F^*}+\ti{e}_{G^{*b}}+\ti{e}_{G^{*r}}\ ,\\[.2cm]
\psi_{\tst2f}(e_s)&=&\ti{e}_{A^*}+\ti{e}_{C^{*b}}+\ti{e}_{C^{*r}}\ ,\\[.2cm]
\psi_{\tst2f}(e_a)&=&\ti{e}_{D^*}+\ti{e}_{F^*}\ .
\end{array}\lb{psif}\ee
The map $\psi_{\tst2f}$ is an embedding. 

Again, one can directly check that the subspace Im$(\psi_{\tst2f})$
is invariant under the operator $M_{\tst2f}$ and the matrix of the induced 
operator in $V_{MS}$, written in the basis (\ref{basms}), coincides 
with $M_{MS}$.

The map $\psi_{\tst2f}$ can be considered as a ``colouring''
of the map $\psi_{gt}$. One can show that this colouring is unique.

\vskip .5cm
4. The map $\psi_{\tst2f}$ also extends to the map from
the five-dimensional space $V_{MS}'$, the second line
in (\ref{psif}) gets replaced by
\be \begin{array}{ccl}
\psi_{\tst2f}(e_m)&=&\ti{e}_{B^*}+2\ti{e}_{F^*}\ ,\\[.2cm]
\psi_{\tst2f}(e_r)&=&\ti{e}_{A^*}+\ti{e}_{G^{*b}}+\ti{e}_{G^{*r}}\ .
\end{array}\ee
However it is still an embedding. 

\vskip .2cm
As we have seen in Subsection 4.1, not only the inflation
matrix but the actual inflation for the Mosseri--Sadoc tiles
(as well as for the five tiles $z$, $m$, $r$, $s$ and $a$)
is induced by the inflation for $\tst2f$.

\section*{Acknowledgments}

Oleg Ogievetsky was supported by the Procope grant 99082.
Zorka Papadopolos was supported by the
Deutsche Forschungsgemeinschaft.
Z. Papadopolos is  grateful for the hospitality to the Center
of the Theoretical Physics in Marseille, where a part of 
this work has been done.
We also thank the Geometry--Center at the
University of Minnesota for making Geomview freely available.

\section*{Appendix: Geodetic angles}

In Section 3.2  we showed that the space of Dehn invariants
for the golden tetrahedra is 2dimensional.
The proof is based on a theorem of Conway, Radin and
Sadun~\cite{CRS}. For completeness we briefly remind
the needed results from~\cite{CRS}.

\vskip .5cm
\ni
{\it Definition.} An angle $\theta$ is called ``pure geodetic"
if $\sin^2\theta$ is rational.

\vskip .5cm
Let ${\cal E}$ be a vector space spanned over $\ku$ by
pure geodetic angles. In~\cite{CRS} a basis of the vector space
${\cal E}$ is constructed. It is useful to know the basis: one
can check whether some given angles are $\ku$--independent.

An element of the basis of ${\cal E}$ is denoted
by $\langle p\rangle_d$. Here $p$ is a prime integer.
The positive integer $d$ has to satisfy two conditions:

\ni 1. $d$ is square--free; 

\ni 2. $(-d)$ is a square modulo $p$.

If $p=2$ then $d\equiv 7\ ({\rm mod}\ 8)$ additionally.

To define $\langle p\rangle_d$ one solves an
equation $4p^s =a^2 + d b^2$ for $a$
and $b$, with a smallest positive $s$. For $d=3$ one requires
$b\equiv 0\ ({\rm mod} \ 2)$; For $d=1$ one requires
$b\equiv 0\ ({\rm mod} \ 4)$.

Now,
\be
\langle p\rangle_d = \frac{1}{s}\ {\rm arccos}
\frac{a}{2p^{s/2}}\ .
\ee

\vskip .5cm
\ni
{\bf Theorem} (Conway--Radin--Sadun). The angles
$\langle p\rangle_d$ together with $\pi$ form a basis
in ${\cal E}$.


\begin{thebibliography}{99}

\bibitem{P}  R.~Penrose, ``The role of aesthetics in pure and
             applied mathematical research'',
             {\em Bull.~Inst.~Math.~Appl.}
             {\bf 10} (1974) 266--271

\bibitem{GS} B.~Gr\"unbaum \& G.C.~Shepard,
             {\em Tilings and Patterns},
             San Francisco: W.H. Freeman 1987

\bibitem{BKSZ} M.~Baake, P.~Kramer, M.~Schlottmann and
               D.~Zeidler,
               ``Planar patterns with fivefold symmetry as
               sections of periodic structures in 4--space"
               {\em Int.~J.~Mod.~Phys.} {\bf B4} (1990)
               2217--2267

\bibitem{K}    P.~Kramer,
               ``Non--periodic Central Space Filling with
               Icosahedral Symmetry using Copies of Seven
               Elementary Cells'',
               {\em Acta~Cryst.} A{\bf 38} (1982) 257--264

\bibitem{SS}   J.~E.~S.~Socolar and P.~J.~Steinhardt,
               ``Quasicrystals. II Unit--cell configurations",
               {\em Phys.~Rev.}~B{\bf 34} (1986) 617--647

\bibitem{MS}   R.~Mosseri and J.~F.~Sadoc,
               ``Two and three dimensional non--periodic
               networks obtained from self--similar tiling'',
               {\em The Physics of Quasicrystals},
               eds. P.~J.~Steinhardt and S.~Ostlund,
               World Scientific (1987) 720--734

\bibitem{D}    L.~Danzer,
               ``Quasiperiodicity; local and global
               aspects",
               in {\em Lecture Notes in Physics 382},
               eds. V.V.~Dodonov and V.I.~Man'ko,
               Springer 1991, pp.~561--572

\bibitem{PHK} Z.~Papadopolos, C.~Hohneker and P.~Kramer,
              ``Tiles--inflation rules for the canonical
              tilings \ts2f \  derived by the projection
              method'', preprint math-ph/9909012
              to be published in the Special Issue of Discrete
              Mathematics in honor of Ludwig Danzer

\bibitem{HKP} C.~Hohneker, P.~Kramer, and Z.~Papadopolos,
              ``Tiles--inflation for the canonical tiling
              $ {\cal T}^{*(2F)}$'',
              in {\em GROUP21 Physical Applications and
              Mathematical Aspects of Geometry, Groups,
              and Algebras, Volume 2},
              eds.: H. -D. Doebner, W. Scherer, and C. Schulte,
              World Scientific, Singapore (1997) 982--986

\bibitem{PO}  Z.~Papadopolos and O.~Ogievetsky,
              ``On Inflation Rules for Mosseri--Sadoc Tilings", 
              preprint, to be published in Proc. of the 
              {\em 7th International Conference on
              Quasicrystals}, Stuttgart, Germany, 
              20-24 September 1999
  

\bibitem{H}    D.~Hilbert,~``Mathematische~Probleme",~{\em  
               Nachr.~Gesellschaft~Wiss. G\"ottingen} 
               (1900)

\bibitem{De}   M. Dehn, ``Uber den Rauminhalt",
               {\em G\"ottingen~Nachr.~Math.~Phys.}
               (1900) 345--354;
               {\em Math.~Ann.} {\bf 55} (1902) 465--478

\bibitem{Sy}   J.~-P.~Sydler,~``Conditions n\'ecessaires et 
               suffisantes
               pour~l'\'equivalence des poly\`edres de l'espace
               euclidien \`a trois dimensions",
               {\em Comm.~Math.~Helv.} {\bf 40} (1965) 43--80

\bibitem{B}    V.~G.~Boltyanski,
               {\em Hilbert's Third Problem},
               Nauka 1977 (in russian)

\bibitem{KPZ}  P.~Kramer, Z.~Papadopolos and D.~Zeidler,
               ``Symmetries of icosahedral quasicrystals'',
               in Proc. of the Symp. {\em Symmetries in
               Science V: Algebraic structures, their
               representations, realizations and physical
               applications}, ed. B.~Gruber et al.,
               Plenum Press (1991) 395--427

\bibitem{GKP}  R. Graham, D. Knuth, O. Patashnik, {\em Concrete
               Mathematics}, Addison--Wesley, 1989

\bibitem{Cl}   A. Clark, {\em Elements of Abstract Algebra},
               Dover Publications, New York, 1971

\bibitem{KP}   P.~Kramer and Z.~Papadopolos,
               ``The Mosseri--Sadoc tiling derived from the
               root--lattice $D_6$",
               {\em Can.~J.~Phys.} {\bf 72} (1994) 408--414

\bibitem{PK}    Z.~Papadopolos and P.~Kramer, 
               ``Models of icosahedral quasicrystals from 6D 
               lattice", {\em Proceedings of the International 
               Conference on Aperiodic Crystals, Aperiodic '94}, 
               ed. by G.\ Chapuis et al., World Scientific, 
               Singapore (1995), pp. 70--76 




\bibitem{CRS}  J.~H.~Conway, C.~Radin, L.~Sadun,
               ``On Angles whose squared trigonometric
               functions are rational'', 
               preprint math--ph/9812019

\end{thebibliography}
\end{document}